\documentclass[runningheads,a4paper]{llncs}
\usepackage{fancyvrb}
\usepackage{graphicx}
\usepackage[utf8]{inputenc}
\usepackage[english]{babel}

\newcommand{\keywords}[1]{\par\addvspace\baselineskip\noindent\keywordname\enspace\ignorespaces#1}

\begin{document}

\mainmatter

\title{Student's opinions about System for automatic assessment of programming tasks Projekt Tomo}
\author{Gregor Jer\v se  \and Matija Lokar}
\institute{Faculty of Mathematics and Physics \\ University of Ljubljana \\ Slovenia
           \and
           Faculty of Computer and Information Science \\ University of Ljubljana \\ Slovenia            
           }
\maketitle

\begin{abstract}
In~\cite{LokarPretnar} a web service called {\em Projekt Tomo} intended to ease the process of 
learning programming for teachers and students has been described. Since the service received a very warm welcome 
from teachers and students alike we decided to collect additional information on the students' 
view of the service in order to improve it even further.

In the paper we briefly present our web service and a detailed analysis of the questionnaire handed out
to the students of the highschool level programming course in Python.
\end{abstract}

\section{Introduction}

Teaching a highschool level programing course can be quite challenging. Since 
programming is a skill, a beginner can only learn it by solving as many 
programming tasks as possible. So teacher has to provide lots of exercises and 
supervise each of the students and help them with their progress.

Since student are beginners they will inevitably make a lot mistakes. If instant 
feedback on what is wrong is not given to them, they can get stuck which slows 
down their progress considerably. This is almost an impossible task for a 
teacher with lots of students and from experience we have observed that a 
considerable amount of students' time is spent just waiting for a teacher to come 
and help them find a mistake in their code. This calls for another approach.

In 2010 we have started the development of a web service \emph{Projekt
Tomo}\footnote{\url{tomo.fmf.uni-lj.si}}. The goal of the web service was to 
support and speed up the learning process. This service was used in teaching 
highschool level programming course for couple of years and we have learned a lot 
while using it. We used the experiences to build the second version of the 
\emph{Projekt Tomo}\footnote{\url{www.projekt-tomo.si}} service in 2015.

The new version has been in use for one year now. To improve the service even 
further we have decided to collect data from users about their opinion on our 
service. That is why we have created a questionnaire and handed it to students. 
The results of the questionnaire are presented in this paper.

\section{Projekt Tomo overview}

This section contains a quick overview of web service Projekt Tomo~\cite{LokarPretnar}. It is freely available online and anyone can login using Facebook, Google or ARNES AAI account. Later one is widely spread in Slovenia, since every student and teacher can receive one. It is Authentication and Authorization Infrastructure for universal network authentication with federated identity.

Once logged in the system, a list of available courses is shown. Each course is divided in problem sets, which contain one or more problems. Every problem is again divided into several tasks. Once a student has found an interesting problem, a file containing a problem description is downloaded to his computer. 

This file contains a problem description and all the necessary code to start solving the problem. A student has to open a file in his favourite IDE and start filling in the solutions. Once a file is executed on a students' computer several things happen. 

First a grading code at the end of the file is executed. The grading code tests students attempts on test cases and reports results to the student. Test cases and the responses can be arbitrarily complex. This step is of crucial importance: it provides instant feedback to the student and works even if the student is offline.

If the web service is reachable the results of the tests together with the source code of the students attempts are submitted to the service and stored. If student later downloads the same problem file his latest attempts are included in the file. The progress of the student can also be seen on the web page of the service. It is meant to motivate the student to solve as many exercises as possible.

The \emph{Projekt Tomo} offers many features that are expected from a service for teaching programming:
\begin{itemize}
    \item low overhead interaction, allowing both the students and the teachers to focus on programming,
    \item remote work, allowing the students to work on the problems anywhere and at any time, while giving the teachers a precise overview of the current progress,
    \item local execution, preventing malicious attacks and allowing the service to have small technical requirements,
    \item independence of both a coding environment and a programming language, allowing the service to be used in various courses,
    \item open source development, allowing the service to be used and expanded by anyone interested in it.
\end{itemize}

\section{Questionnaire}

The main goals when designing Projekt Tomo web service was to provide easy-to-use web service for learning programming that provides instant feedback to the users. To see if our goal has been achieved and to improve our service even further we have decided to collect information directly from the users of the web service. Thay is why we have created a questionnaire and handed it out to the students who have used the service to complete a highschool-level programming course. Most of the students have solved at least 50 problems using the web service so we can assume that they are familiar with the system and have been using it regularly. The answers to the questionnaire have been submitted anonymously by more than 50 students.

\subsection{Ease of use}

Students have been asked about their impression about the ease of use, speed, usability and design of the service. The ease of use was graded on the scale from 1 to 5. More than 90 percent of the students responded with grade 4 or 5. The the average student grade was 4.4 with standard deviation 0.83. More 70 percent of the students were also very satisfied with the speed, usability and the design of the service. So our goal to create easy-to-use web service has definitely been achieved.

\subsection{Impression about the service}

We have asked students to grade the importance of offline work, instant feedback and the possibility to inspect the official solution. The possibility of offline work was very important to about half of the students. But the instant feedback and the possibility to inspect the official solution were of great importance to 84 percent of the students. This assures us that our design decisions were correct and main goals reached.

The most valuable part of this survey was a text entry box where students expressed their overall impression about the service. All but one of the comments were very positive. In the summary they praised the design, usability, ease-of-use and the ability to get instant feedback. Some of the typical responses were:
\begin{itemize}
 \item top web service for teaching programming,
 \item simple-to-use, responsive,
 \item I like that it checks the solutions and provides instant feedback.
\end{itemize}

The only negative feedback mentioned speed problems when saving solutions to the web service. The issue was investigated and attributed to the network configuration at the school.

Most of the students would also recommend the service to their school mates, mostly because of the ability to get instant feedback and ease of use. Based on their opinion we have decided to spread the use of the service to a larger scale. 
We have already started a project~\cite{CS4HS} whose partial goal is also to create a repository of exercises that will be used to teach computer science in Slovenian high-schools based on the new E-textbook~\cite{ucbenik}.

\subsection{What students liked and disliked}

Students specified three things they most liked about the service. The overall winner of this section was definitely instant feedback, which was mentioned in almost all responses. As before, students have also praised usability, easy-to-use and design. 

Some of them have also mentioned that the instant feedback and easy progress tracking on the web page motivates them to solve more programming exercises than they would without the service. For instance one student wrote 

``the service looks like an survey or a "computer game" where each solved exercise increases your level of knowledge. You are getting feedback what you are doing right and what not. All successfully solved exercises are colored green. What I haven't solved is colored orange or red so I can instantly see what exactly is missing, so I can return to the specific exercise later.''

They were also asked to specify three things they did not like about the service. The highlight of the comments was bad instant feedback. This can be solved by educating teachers about how to produce quality exercises for our service.

\subsection{Suggestions}

We also got some suggestions on how to improve the web service. Their suggestions were highly correlated with the things the students did not like. Most of them would improve the test cases to provide more detailed instant feedback about what is wrong about their solution. 

One student suggested to add support for programming languages other than Python and one suggested to add the possibility to make official solution visible after some unsuccessful attempts have been made.

\section{Future extensions}

Our goal is to create the best web service for teaching programming. In order to do that, we have to constantly improve our service. Based on the information obtained in the questionnaire we have decided to invest our time in improvement of the quality of exercises and to make writing quality test cases for exercises even easier, so it will be easier for teachers to produce quality exercises for the service. 

Some students have mentioned that it is very frustrating when they unsuccessfully try to solve an exercise several times. They suggested that the ability to show the official solution after some number of unsuccessful attempts have been made would be helpful. Based on this comment we have decided to implement this possibility.

Currently only supported programming languages are Python and Octave. Based on the provided improvement suggestions our next goal will be to add support for more programming languages to the service in the near future. 

\section{Conclusions and future work}

As the pilot study of web service usage has been quite successful, Projekt Tomo web service has been chosen as one of the cornerstones of the project within  
Google's Computer Science for High School~\cite{CS4HS} awards program for 2016. The project's goal is to develop class-ready resources for CS teachers, especially for topics connected with programming and algorithms. Namely in first reactions to our new e-textbook for CS, which was introduced in 2015~\cite{ucbenik}, teachers reported lack of knowledge in the  domain of programming and algorithms and asked for assistance. TOMO platform will be used for exercises. So it is expected that the usage of TOMO will be significantly increased in the forthcoming  years. We intend to closely follow all reactions, responses and suggestions and also to conduct the study on usefulness of using this approach in learning programming.

\end{document}